\documentclass[a4paper,11pt]{article}
\pdfoutput=1 % if your are submitting a pdflatex (i.e. if you have
\usepackage{jinstpub} % for details on the use of the package, please
\usepackage{siunitx} % FG for si unit
\DeclareSIUnit\sq{\ensuremath{\Box}}                           % \sq print a square symbol

\usepackage[rawfloats=true]{floatrow}

\usepackage{subcaption}
\makeatletter
\renewcommand\p@subfigure{\thefigure.}                        % ref of subcaption is X.Y
\makeatother

\title{CACTUS: A depleted monolithic active timing sensor using a CMOS radiation hard technology}

\author[a,1]{Y.~Degerli\note{Corresponding author.}}
\author[a]{, F.~Guilloux}
\author[a]{, C.~Guyot}
\author[a]{, JP.~Meyer}
\author[a]{, A.~Ouraou}
\author[a]{, P.~Schwemling}

\author[b]{, A.~Apresyan}
\author[b]{, R.~Heller}
\author[b]{, M.~Mohd}
\author[b]{, C.~Pena}

\author[c]{, S.~Xie}

\author[d]{, T.~Hemperek}

\affiliation[a]{IRFU, CEA, Universit\'{e} Paris-Saclay,\\F-91191 Gif-sur-Yvette, France}
\affiliation[b]{Fermi National Accelerator Laboratory\\ Wilson Street and Kirk Road, Batavia, 60510, IL, USA}
\affiliation[c]{California Institute of Technology\\ 1200 E California Blvd, Pasadena, CA 91125, USA}
\affiliation[d]{Physikalisches Institute der Universit\"{a}t Bonn\\Nussallee 12 D-53115. Bonn, Germany}

\emailAdd{yavuz.degerli@cea.fr}

\abstract{
The planned luminosity increase at the Large Hadron Collider in the coming years has triggered interest in the use of the particles' time of arrival as additional information in specialized detectors to mitigate the impact of pile-up. The required
time resolution is of the order of tens of picoseconds, with a spatial granularity
of the order of \SI{1}{mm}. A time measurement at this precision level will also be of interest beyond the LHC and beyond high energy particle physics. We present in this
paper the first developments towards a radiation hard Depleted Monolithic
Active Pixel Sensor (DMAPS), with high-resolution time measurement capability.
The technology chosen is a standard high voltage CMOS process, in conjunction with a high resistivity detector material, which has already proven to efficiently detect particles in tracking applications after several hundred of Mrad of irradiation.
}


\arxivnumber{2003.04102} % only if you have one

\begin{document}
\maketitle
\flushbottom

\section{Introduction}
\label{sec:intro}
Time measurements with a resolution of \SI{100}{ps} or better are rapidly gaining
interest in detector development for high energy physics. Such a time
resolution on individual Minimum Ionizing Particles (MIP) gives the potential
to associate individual particles to their production vertex in high pile-up 
environments like the high luminosity upgrade of the Large Hadron Collider (HL--LHC) with up to 
200 average interactions per bunch crossing.

Several timing detector projects within the ATLAS and CMS experiments are being
developed right now, with the aim of resolutions of the order of \SI{30}{ps} per
MIP. Prominent examples include the High Granularity Timing Detector (HGTD) in ATLAS~\cite{Allaire1} and the MIP Timing Detector (MTD) in CMS~\cite{CMS:2667167,APRESYAN2018158}, which both use LGAD (low-gain avalanche diode) technology.
% FG : I propose to remove this part

%We present in this paper the design, simulation and test results of a possible
%long term alternative to LGADs, based on a 
We present in this paper the design, simulation and test results of an integrated timing detector based on a 
Depleted Monolithic Active
Pixel Sensor (DMAPS), using an industrial \SI{150}{nm} CMOS process from LFoundry with High Resistivity (HR) wafers. The choice of a CMOS process is based on the three main benefits expected from standard DMAPS technology: low cost, radiation hardness, and good time resolution.

High Voltage (HV) CMOS processes are relatively low cost since they are the process of choice for high volume consumer electronics used
in the automotive industry. In addition, a monolithic architecture eliminates the need for costly bump bonding operation~\cite{Pieric1}. 
Altogether, the DMAPS opens the possibility for relatively cheap sensors, which is crucial for future large projects such as the FCC. HV--HR CMOS technologies such as the LFoundry \SI{150}{nm} LF15A have proven to accept a high bias voltage of the order of \SI{300}{V} applied on the sensor backside, over a thickness of \SI{200}{\micro\m}, fully depleting several hundred micrometers of silicon material. They have been tested under irradiation for tracking applications, and a suitable signal to noise ratio has been measured for doses up to \SI{150}{Mrad}~\cite{Chen1,Barbero1}. Recently, dedicated monolithic sensors developed in a SiGe BiCMOS process for a Time-Of-Flight PET application have proven a timing resolution of the order of \SI{50}{ps}~\cite{Iacobucci1}. This demonstrates the high potential of CMOS processes for timing applications.

\section{Architecture of the sensor}
\label{sec:architecture}
\subsection{Theoretical timing resolution limits}
Reference~\cite{Riegler_2017} gives analytical expressions for the time resolution of silicon detectors without intrinsic signal multiplication. The charge deposit of high energy particles is described with the Landau theory. The three components contributing to the
time resolution, namely the charge deposit fluctuations, noise and fluctuations of the signal shape are all estimated analytically. The effect of leading edge discrimination is also taken into account. The main conclusions are :
\begin{itemize}
%\item The thinner the sensor, the better
%the time resolution. Assuming \SI{200}{V} applied to a sensor with thickness of \SI{200}{\micro m}, the time resolution should be \SI{180}{ps},
%and \SI{59}{ps} of a \SI{100}{\micro m} thick sensor.
\item For MIP detection, the thinner the sensor, the better
the time resolution. Assuming the sensor is biased at \SI{200}{V}, the time resolution should be \SI{180}{ps} and \SI{60}{ps} for a sensor thickness of \SI{200}{\micro m} and \SI{100}{\micro m} respectively.
\item To not degrade the time resolution using leading edge discrimination, the peaking time should be rather large (of
the order of \SI{1}{ns}), and the threshold set at low values, typically below \SI{40}{\%} of the total signal charge.
\end{itemize}
The main parameters of the sensor discussed in this work are close to these settings in terms of depletion depth, electric field  and front end architecture.   

\subsection{Pixel Design}
The pixel design of the CACTUS (Cmos ACtive pixel Timing \Large{\si{\micro}} \normalsize Sensor) chip presented in this paper is largely based on previous pixel designs, with many
modifications to improve the timing performance. The pixel designs we
started from are two tracking demonstrator chips, LF-CPIX~\cite{Degerli1,Degerli2} and LF-MONOPIX~\cite{Wang1}, originally intended for the
$5^{th}$ layer of the ITK, the upgrade of the ATLAS internal tracker for the HL-LHC. The design technology used for these demonstrators is the same as for the CACTUS, and uses the LFoundry \SI{150}{nm} LF15A process, with high resistivity substrate (\SI{>= 2}{k \ohm .cm)}.

Compared to LF-CPIX and LF-MONOPIX, we have chosen to implement large surface collecting diodes of \SI[product-units = power]{1 x 1}{\mm} and \SI[product-units = power]{1 x 0.5}{\mm}. There are several reasons to use such large collecting diodes
for a timing detector:
\begin{itemize}
\item A pitch of about \SI{1}{mm} is the order of magnitude of what is intended
for detector projects like the HGTD and is representative of the required pitch
for future timing oriented detectors.
\item To speed up the output signal and reduce as much as possible its rise time, it is needed to increase significantly the bias currents of the front-end electronics. Small pixel pitches would lead to an
unacceptable power dissipation per surface unit. In our case, for an estimated bias current of \SI{800}{\micro A} per pixel, the static power consumption of the active area is \SI{145}{mW/cm^2}.
\item To ensure the best possible timing resolution, the electric field lines within the collection diode have to be as parallel as possible. A large collection diode allows reducing the fraction of the surface where fringe effects are significant, and large pads help to minimise the influence of shared hits.
\end{itemize}

\begin{figure}[h]
\centering
  \floatsetup{heightadjust=all, valign=c}
  \begin{floatrow}
  \ffigbox{%
    \includegraphics[width=0.45\textwidth, height=5.2cm, keepaspectratio]{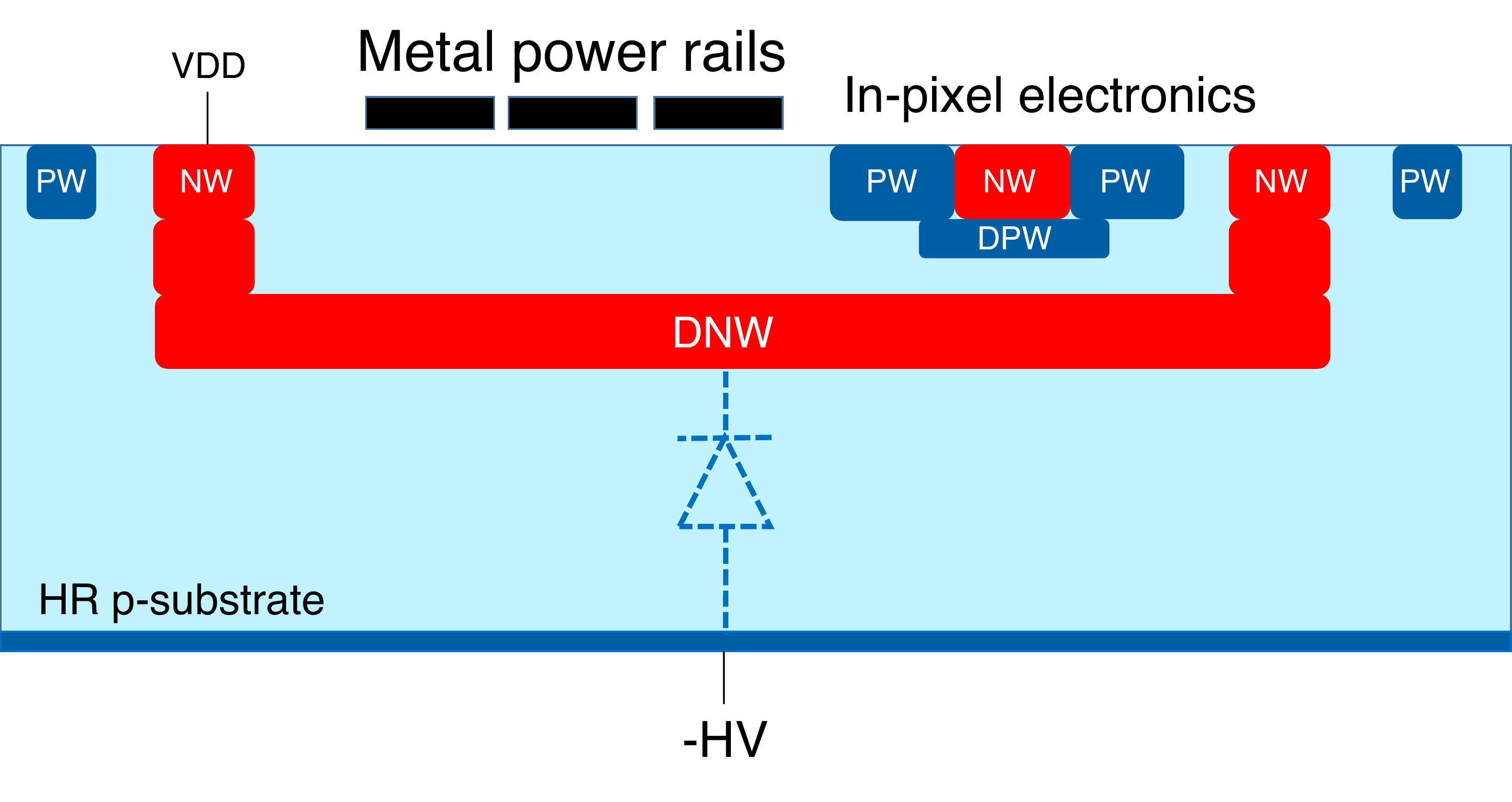}} {
    \caption{Cross-section of a pixel showing the in-pixel electronics and the metal power rails (drawing is not to scale).}
    \label{fig:cross-section}}
  \ffigbox{%
    \includegraphics[width=0.45\textwidth, height=5.2cm, keepaspectratio]{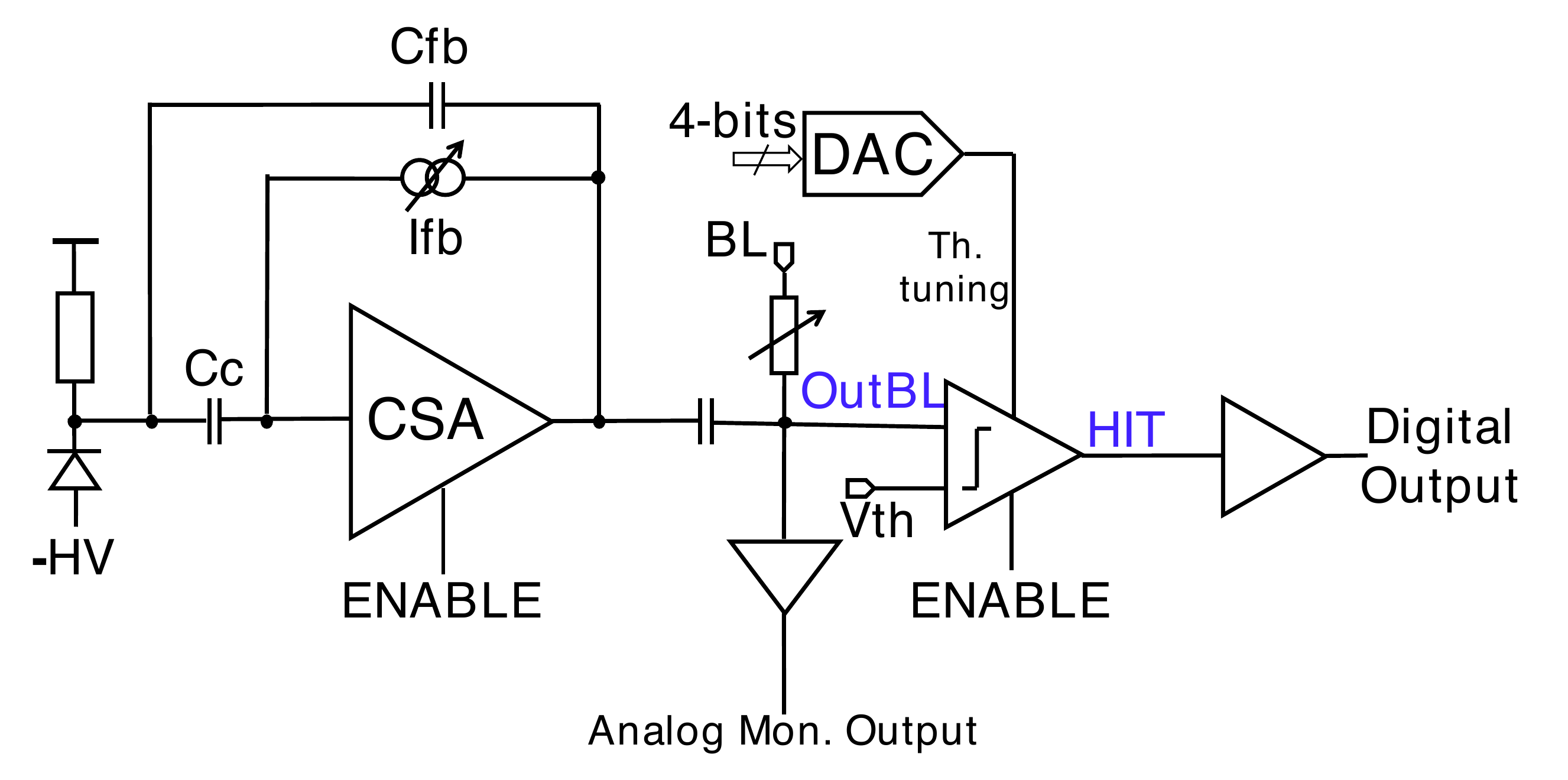}} {
    \caption{Block diagram of the CACTUS pixel analog front-end.}
    \label{fig:schematics}}
  \end{floatrow}
 \end{figure}

The cross-section of a pixel of CACTUS is shown in figure~\ref{fig:cross-section} (not to scale). Due to the deep n-well (DNW) and buried p-well (DPW) layers available in this process, it is possible to implement complementary (NMOS and PMOS) transistors inside the pixel. The in-pixel electronics occupy only a small part of the area of the pixel. On the other hand, the power rails are quite large to minimize the resistivity of the supplies. 

% FG : Single figure code kept for easy fall back
%\begin{figure}[h]
%\center{\includegraphics[width=.6\linewidth]{images/pixel.png}}
%\caption{Cross-section of a pixel showing the in-pixel electronics and metal power rails (not to scale).}
%\label{cross-section}
%\end{figure}

% FG : test of 2 figures side by side

The in-pixel front-end electronics of each pixel is made of a fast charge-sensitive amplifier (CSA) followed by a leading-edge discriminator (see figure~\ref{fig:schematics}). Each pixel has a 4-bit DAC to correct any threshold non-uniformity from the discriminators. The analog output of the CSA (OutBL) and the digital output of the discriminator (HIT) can be monitored out of the chip through dedicated on-chip buffers. The charge collection diode is continuously biased through a very
high impedance device (a diode-connected NMOS transistor~\cite{Degerli1}). There is no leakage compensation circuit.

To ensure good timing resolution with such
a configuration, the time walk effects have to be measured and corrected. If the signal shape has sufficient uniformity over the sensitive surface, a
unique time-walk correction can be used independent of the impact point of the particle.

% FG : Single figure code kept for easy fall back
%\begin{figure}[h]
%\center{\includegraphics[width=.6\linewidth]{images/schematics.png}}
%\caption{Principle schematics of the CACTUS pixel analog front-end}
%\label{schematics}
%\end{figure}

The implementation of two pixel sizes allows 
to study the front-end performance for two detector capacitance values. The pixel capacitance values have been estimated from the Sentaurus TCAD simulation software ~\cite{Synopsys} and the Cadence Integrated Circuit (IC) design tools by simulating the contribution of the diode and the in-pixel electronics. The capacitance values are around \SI{1.5}{pF} and \SI{1}{pF} for pixel sizes of \SI{1}{\mm^2} and
\SI{0.5}{\mm^2} respectively. Other critical parameters concerning the detection
diode have also been checked using Sentaurus software: the breakdown voltage,
the total charge collected, and the charge collection as a function of time. Typically, for a \SI{100}{\micro m} thin sensor, according to these simulations almost all of the charge generated from a MIP is collected within \SI{5}{ns},
with a rise time of the order of \SI{1}{ns}~\cite{Guilloux1}.

\begin{figure}[h!]
\center{\includegraphics[width=0.8\linewidth, height=5.2cm, keepaspectratio]{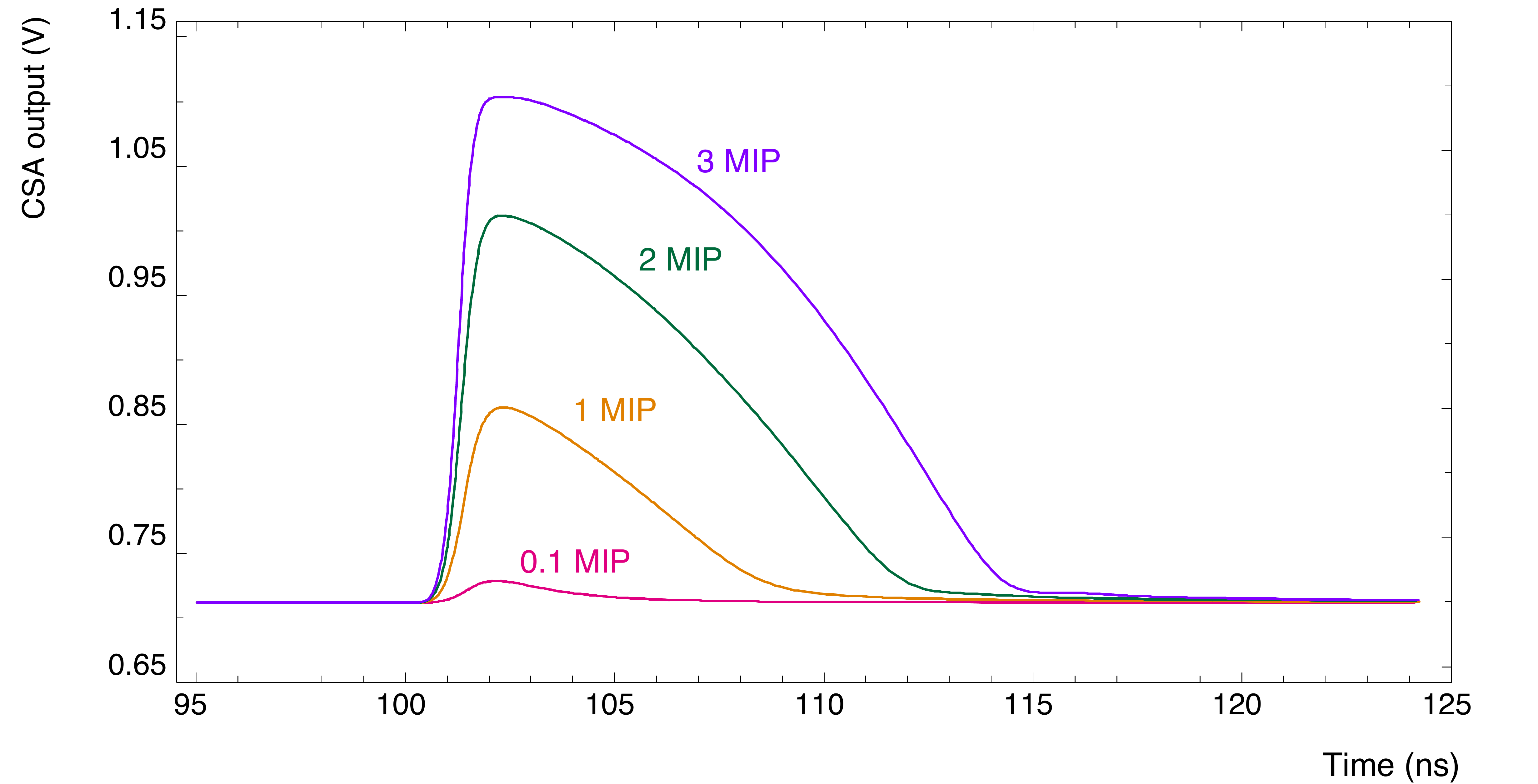}}
\caption{Cadence transient simulation result of the analog front-end for a \SI{100}{\micro m} thinned  sensor. The input charge varies from \SI{1}{ke^-} (0.1 MIP ) to \SI{24}{ke^-} (3 MIP). The front end rise time is about \SI{1}{ns} (VDD~$=\SI{1.8}{V}$, Ibias~$=\SI{800}{\micro A}$).}
\label{simulation-cadence}
\end{figure}

The front-end performance has been studied using Cadence simulations tools. Transient simulation results with an injected charge varying between \SI{1}{ke^-} and \SI{24}{ke^-} (corresponding to about \SI{0.1}{MIP} to \SI{3}{MIP} for a \SI{100}{\micro m} thinned sensor) are shown in figure~\ref{simulation-cadence} for a detection capacitance of \SI{1.5}{pF} (VDD~$=\SI{1.8}{V}$, total bias current~$=\SI{800}{\micro A}$). According to these simulations, the rise time is about \SI{1}{ns} and the input-referred noise \SI{300}{e^-}.

\subsection{Chip Design}
\begin{figure}[h!]
\center{\includegraphics[width=.5\linewidth, height=5.2cm, keepaspectratio]{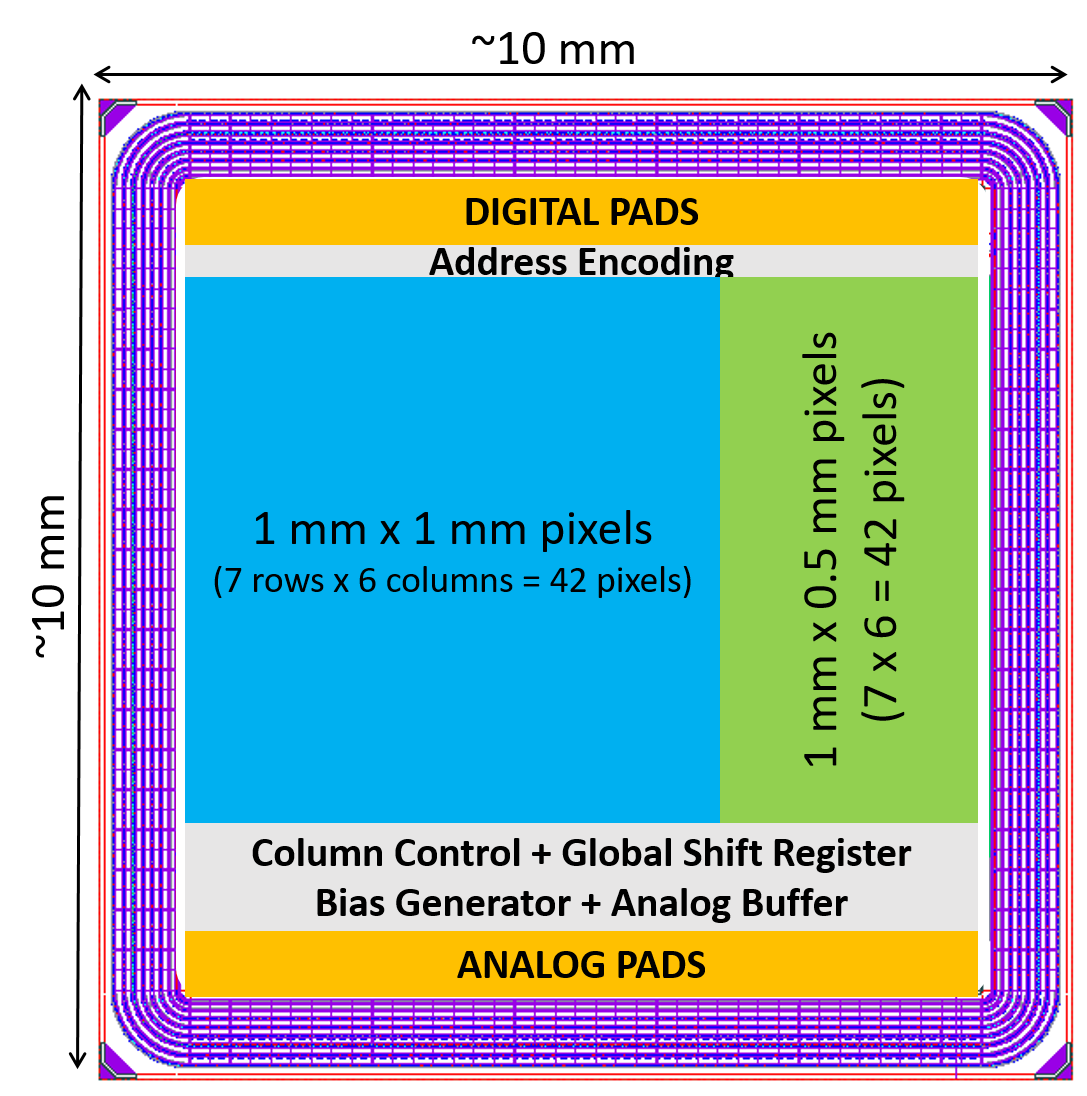}}
\caption{Simplified block diagram of CACTUS. The external guard ring width is \SI{250}{\micro\m} for CACTUS versions~A and B.}
\label{cactus-layout}
\end{figure}

The global layout of the chip is shown in figure~\ref{cactus-layout}. The total chip surface
is about \SI{100}{mm^2}. In addition to the detection pixels with their front-end
electronics, the chip features a slow control system allowing to enable or disable
any individual pixels and to set the bias currents of global DACs. The control data are loaded
using a SPI-like bus implementation and stored on-chip in a shift register of \SI{281}{bits}. The chip includes a column-encoding logic, giving the coordinate of the hit pixel on dedicated output channels. 

For each column, the pixels share a common readout line which sends the discriminator signals to a fast output (LVDS and/or CMOS). There is
also a global digital output signal (HITOR) and an analog monitoring output that can
be connected to any of the pixels (only 1 chosen pixel at a time). The analog monitoring buffer, which is a cascade of 2 distributed source followers, has a limited bandwidth compare to the in-pixel CSA. Its gain is around $0.6$.

Two versions of the chip have been designed and submitted to fabrication. The main difference between them is the external guard-rings used to apply the HV on the substrate. While the version~A uses the guard-rings from LF-CPIX (V2) with a proven breakdown voltage (BV) of about \SI{220}{V}~\cite{Degerli2}, the version~B uses an improved guard-ring strategy. The expected BV voltage for the version~B is greater than \SI{350}{V}. The width of the guard-rings for both versions is \SI{250}{\micro\m}.

\section{In-lab characterisation}
\label{sec:labtests}
After the fabrication of CACTUS sensors, one wafer has been diced without post-processing. The thickness of a standard wafer is \SI{725}{\micro\m}. 
The increase of depletion depth increases the total collected charge but increases also the charge collection time and the
fluctuations on the charge deposition. An optimum should be found for this critical parameter. Two additional wafers have been thinned to \SI{200}{\micro\m} and \SI{100}{\micro\m} and then post-processed with boron implant and metallization, to allow back-side polarization of the sensor. Some \SI{100}{\micro\m} chips sensors have been tested in lab, and \SI{200}{\micro\m} sensors have been characterized in-lab and in test-beam.

%In order to study the effect of sensor thickness, 2 other wafers have been thinned to \SI{200}{\micro\m} and \SI{100}{\micro\m} and then post-processed for back-side polarization of the sensor. 

For laboratory tests of the sensor, a dedicated test-bench has been developed. A proximity board (called the "CACTUS board") holds the CACTUS sensor and some drivers. This board is connected to a General-purpose Programmable Analog Card (GPAC) which provides the low voltage power supplies, references, and buffering of fast CACTUS signals. A Raspberry-Pi connected to the GPAC via a small adapter board is used to generate the slow control signals and also to read out the addresses of the hit pixels. % A WaveCatcher, an 8 channel 12 bit switched capacitor array based fast digitizer~\cite{WC}, is used for fast sampling of analog and digital outputs.
A WaveCatcher, an 8 channel 12 bit fast digitizer~\cite{WC} based on a switched capacitor array, is used for fast sampling of analog and digital outputs.

\subsection{I-V measurements}
\label{subsec:IV}
\begin{figure}[h]
\center{\includegraphics[width=\linewidth, height=5.2cm, keepaspectratio]{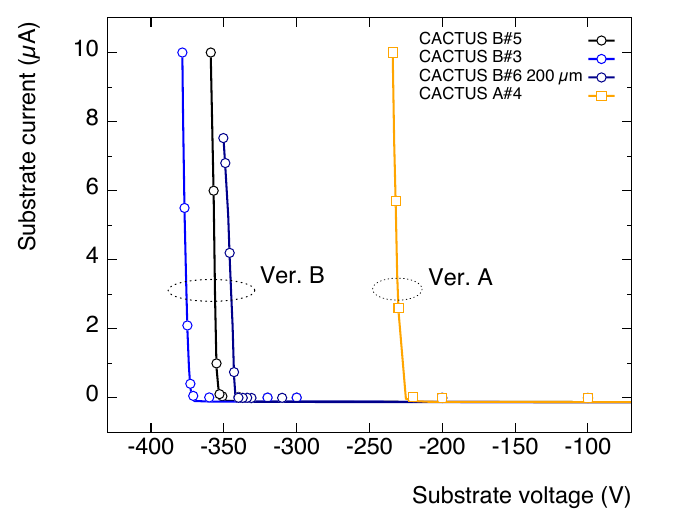}}
\caption{Break-down voltages measured on different \mbox{CACTUS} chips (versions A and B).}
\label{fig:bv}
\end{figure}

The BVs of the charge collection diodes have been evaluated by increasing the substrate bias voltage of the chip while measuring the supply current. Figure~\ref{fig:bv} shows the BVs measured on four different non-thinned and thinned CACTUS chips. The BV measured on version~A is \SI{230}{V}. The BV measured on version~B is around \SI{350}{V}. The value for version~A was also confirmed on other thinned and non-thinned sensors. According to previous tests with similar structures and the same substrate resistivities on the same process, these voltages are enough to fully deplete thinned substrates (\SI{-15}{V} needed for \SI{100}{\micro m} and \SI{-60}{V} for \SI{200}{\micro m}~\cite{Caicedo1}). The ability to bias the sensor with higher voltages is beneficial to compensate for the charge losses after neutron irradiation~\cite{Mandic1}.

\subsection{Signal to noise characterisation}
\label{subsec:S2N}
\begin{figure}[h]
\begin{subfigure}[b]{0.49\linewidth}
\center \includegraphics[width=\linewidth]{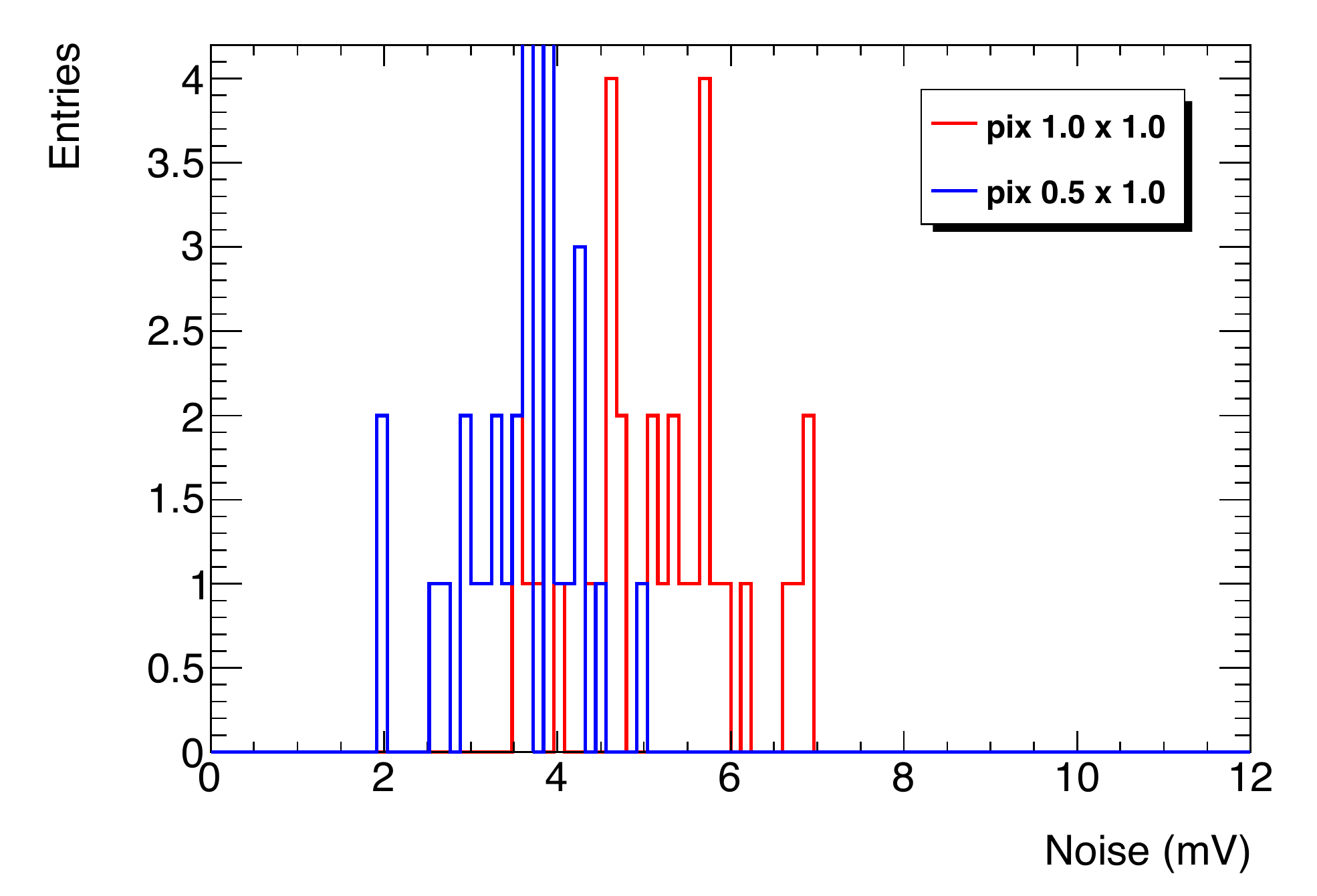}
\caption{}
\label{subfig-noise-threshold-a}
\end{subfigure}
\begin{subfigure}[b]{0.49\linewidth}
\center \includegraphics[width=\linewidth]{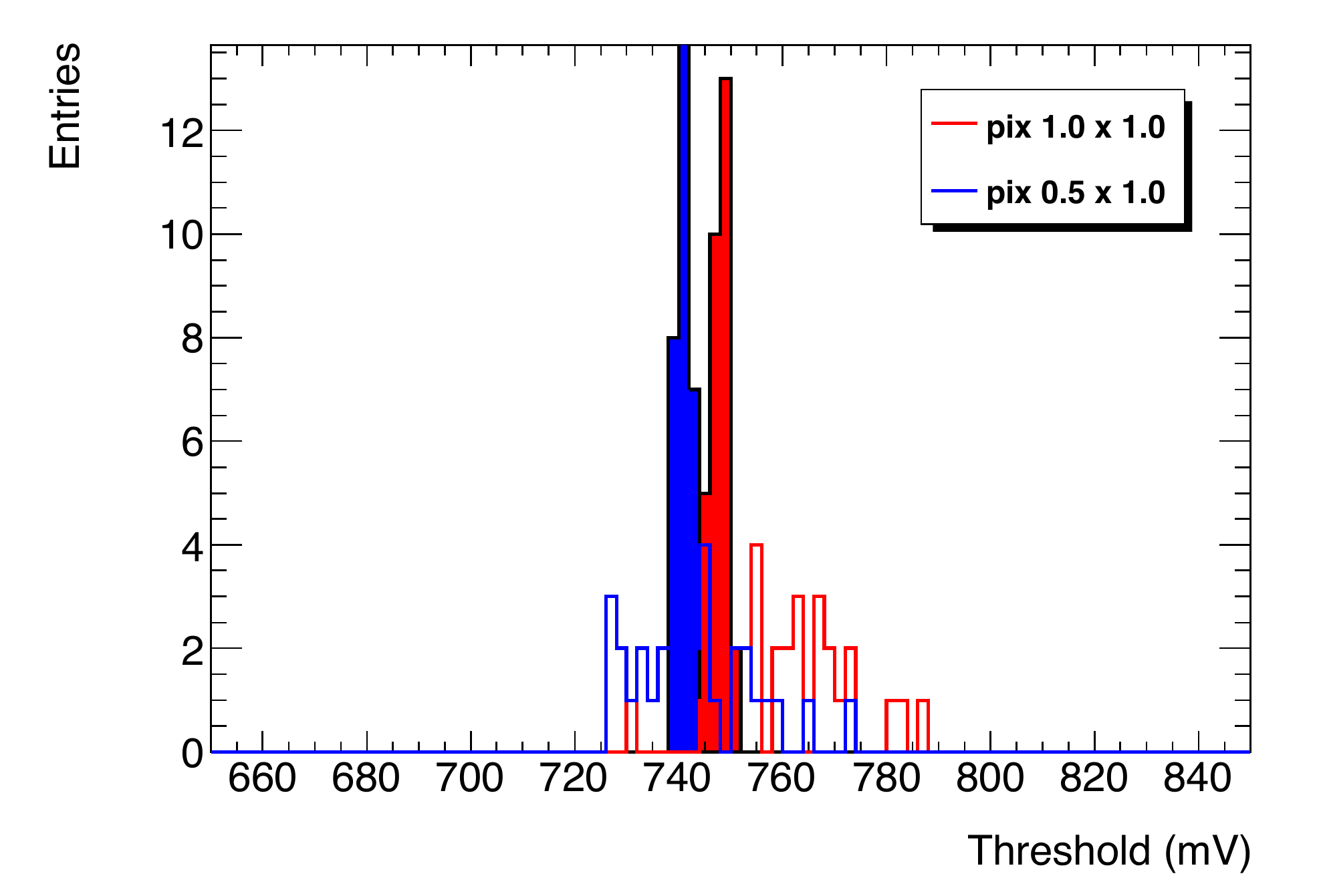}
\caption{}
\label{subfig-noise-threshold-b}
\end{subfigure}
\caption{a) Temporal noise and b) threshold distributions of all pixels of a CACTUS ver.~B before (wide distribution) and after tuning (narrow distribution).} 
\label{fig:noise-threshold-bis}
\end{figure}

The temporal noise and threshold (off-set) dispersions have been evaluated for each pixel from \mbox{S-curves} obtained by scanning the external threshold voltage (Vth in figure~\ref{fig:schematics}). Figure~\ref{fig:noise-threshold-bis} shows the distributions of these parameters obtained from a non thinned sensor (version B) with VDD~$=$~\SI{1.8}{V}, VBL~$=\SI{700}{mV}$, a total bias current of \SI{800}{\micro A}, and an HV~$=\SI{0}{V}$. From   figure~\ref{subfig-noise-threshold-a}, the mean RMS noise is \SI{3.56}{mV} for small and \SI{5.20}{mV} for large pixels (this should correspond to \SI{230}{e^-} and \SI{300}{e^-} input-referred noise respectively according to initial simulations). The threshold dispersions are high before tuning (large distributions in figure~\ref{subfig-noise-threshold-b}), but after tuning to an arbitrary value of \SI{740}{mV}, due to in-pixel DACs, they become lower than the temporal noise values: \SI{1.46}{mV_{RMS}} for small pixels and \SI{1.68}{mV_{RMS}} for large pixels (corresponding to \SI{94}{e^-} and \SI{96}{e^-} respectively according to simulations). The mean values are \SI{747}{mV} before tuning and \SI{740}{mV} after tuning (corresponding to $\simeq \SI{3100}{e^-}$ and $\simeq \SI{2600}{e^-}$ respectively according to simulations). The results are the same when HV~$=\SI{-300}{V}$. Similar tests done on thinned sensors show that thinning and post-processing do not change the electrical characteristics of the front-end. This is also valid for the high voltage sensor bias.

The signal to noise measurements have been done using a radioactive $^{133}Ba$ source, whose main emission line is at \SI{31}{keV} (corresponds to $\simeq \SI{8.5}{ke^-}$ in silicon). The ASIC (application-specific integrated circuit) tested is a non-thinned version B sensor, biased at HV~$=\SI{-300}{V}$.

\begin{figure}[h]
\center{\includegraphics[width=\linewidth, height=5.2cm, keepaspectratio]{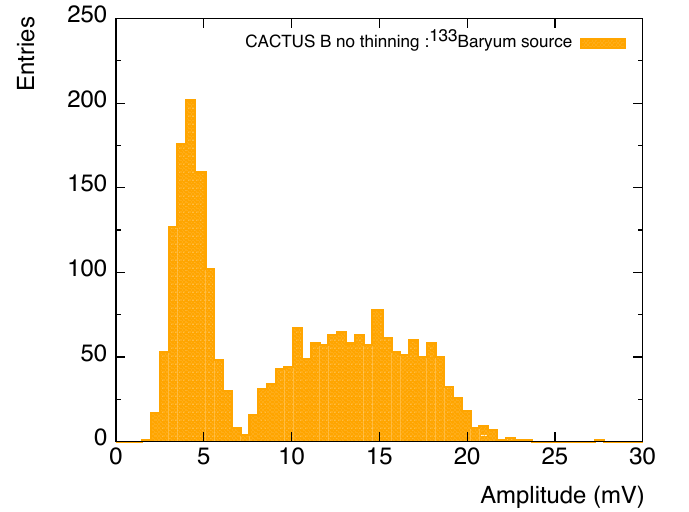}}
\caption{Spectrum from a $^{133}Ba$ source obtained at the analog monitoring output of a \SI[product-units = power]{1 x 0.5}{\mm} pixel. The noise and the signal are clearly separated. The MPV of the signal is measured around \SI{14}{mV}, six times lower than simulated.}
\label{fig:baryum}
\end{figure}

Figure~\ref{fig:baryum} shows the spectrum obtained for one \SI[product-units = power]{1 x 0.5}{\mm} pixel at the analog monitoring output. The signal amplitude measured is almost six times lower than expected from simulations (measured MPV around \SI{14}{mV} instead of \SI{96}{mV}). The source of the reduced gain was found to be an underestimated extraction of the in-pixel power rail capacitances with the extraction software tools (capacitances between metal layers and the large HR p-type area inside the DNW of the pixel, shown in  figure~\ref{fig:cross-section}). New estimations of the actual capacitance lead us to the conclusion that the input diode capacitance is more than an order of magnitude larger than expected (more than \SI{15}{pF}, see simulation results in table~\ref{tab:simuAmp}). The open loop gain of the charge sensitive amplifier is not large enough to compensate for the very high input capacitance value and the overall gain of the pixel is highly degraded.
%One of the consequences of this gain problem is the increase of noise.
If the input diode capacitance is around \SI{15}{pF} then the input referred temporal noise levels should be of the order of \SI{2}{ke^-} instead of \SI{230}{e^-}.

\begin{table}[h]
    \centering
    \caption{Simulated analog monitoring output amplitude as a function of CSA input capacitance (\SI{8.5}{ke^-} input charge).}
    \begin{tabular}{|S|S|}
         \hline
         {$C_{det}$ (pF)} &  {Signal amplitude (mV)}\\ \hline
           1.0            & 96.4    \\ 
           1.5            & 88.8    \\ 
          15.0            & 17.8    \\ 
          20.0            & 13.4    \\           
         \hline        
    \end{tabular}
    \label{tab:simuAmp}
\end{table}

\subsection{Timing measurements with $^{90}Sr$ source}
\label{subsec:SR}
For timing measurements, a collimated $^{90}Sr$ source is placed on top of the Device Under Test (DUT) at a distance of \SI{8}{cm}. A photomultiplier (PMT) (Hamamatsu 11934) coupled to a plastic scintillator and placed below the DUT provides the reference time measurement. Both CACTUS digital outputs and PMT signals are sent to a WaveCatcher digitizer~\cite{WC}.

The data coming both from the Raspberry-Pi (the address of the hit cell and the trigger) and from the WaveCatcher (CACTUS signals) are collected on a computer. The arrival time difference of the PMT and CACTUS signal is used to estimate the CACTUS timing resolution. Independently the PMT resolution has been measured around \SI{80}{ps} using two PMTs and a $^{22}Na$ source. 

\begin{figure}[h]
\begin{subfigure}[b]{0.5\linewidth}
\center \includegraphics[width=\linewidth]{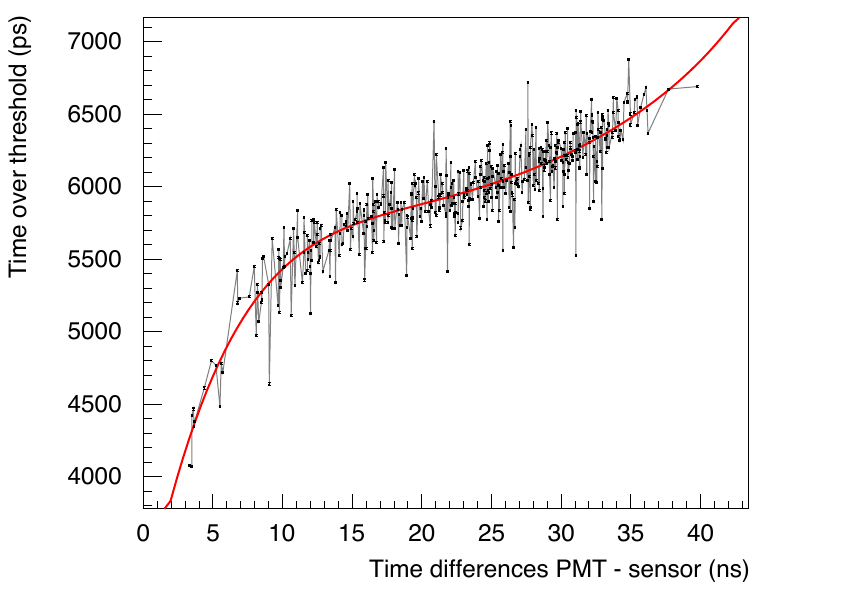}
\caption{}
\label{subfig_Sr90-a}
\end{subfigure}
\begin{subfigure}[b]{0.5\linewidth}
\center \includegraphics[width=\linewidth]{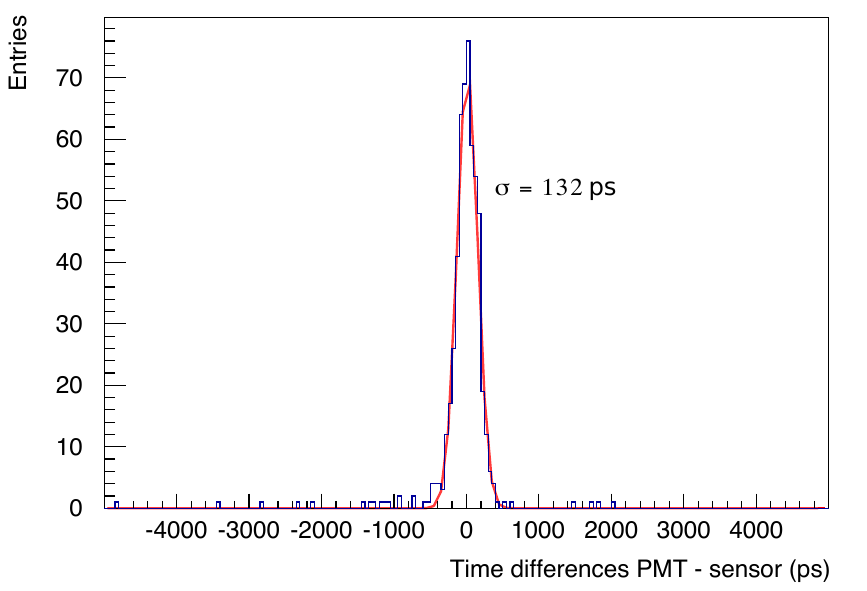}
\caption{}
\label{subfig_Sr90-b}
\end{subfigure}
\caption{Time differences between sensor and PMT a) versus TOT and polynomial parameterization. b) After TOT correlation corrections. The \SI{132} {ps} dispersion corresponds to the quadratic sum of the PMT \SI{80} {ps} resolution and the CACTUS \SI{105} {ps} resolution (threshold~$\simeq \SI{2.9}{MIPs}$).}
\label{Sr-90}
\end{figure}

The CACTUS used in these measurements is a \SI{200}{\micro m} thick sensor biased at HV~$=\SI{-300}{V}$.
At the price of a high trigger threshold (\SI{80}{mV} corresponding to $\simeq \SI{2.9}{MIPs}$), using an ad hoc TOT correction (figure~\ref{subfig_Sr90-a}) we obtained a CACTUS time resolution of \SI{105}{ps} (figure~\ref{subfig_Sr90-b}). The TOT correction accounts for the PMT timewalk, the CACTUS timewalk and the correlations between time of flight and energy deposit. The result shown in figure~\ref{subfig_Sr90-b} includes the contribution of the PMT time resolution. On figure~\ref{subfig_Sr90-a}, the highest amplitude energy deposits correspond to the slowest $\beta$ particles, and hence to the highest time difference between the PMT and the sensor, and also to the highest PMT TOT. At lower thresholds, the timing resolution degrades: \SI{278}{ps} at \SI{1.7}{MIPs} and \SI{303}{ps} at \SI{1.0}{MIPs}.

\section{Test beam}
\label{sec:testbeam}

A test beam campaign has confirmed the signal to noise ratio estimated from X-ray sources and allowed us to study the uniformity and charge collection of the sensor. It also gave an estimation of the time resolution for MIPs.

\subsection{Test beam infrastructure}
\label{sec:testbeaminfrastructure}
Test beam measurements were performed at the Fermilab Test Beam Facility (FTBF)~\cite{FTBF}, 
which provides a unique opportunity to characterize prototype detectors for collider
experiments. A typical application is to place a DUT
in the high energy beam, and measure its response to the beam particles passing through its active area. FTBF provides a \SI{120}{\GeV}
proton beam from the Fermilab Main Injector accelerator. The FTBF beam 
is resonantly extracted in a slow spill for each Main Injector cycle 
delivering a single \SI{4.2}{sec} long spill per minute, tuned to yield approximately
\SI{60000}{protons} per single spill. The primary beam (bunched at \SI{53}{MHz}) consists 
of \SI{120}{\GeV} protons. All measurements presented in this paper were taken with
the primary beam particles.

\begin{figure}[h!]
  \begin{subfigure}[b]{0.55\linewidth}
    \center \includegraphics[width=\linewidth, height=6.5cm, keepaspectratio]{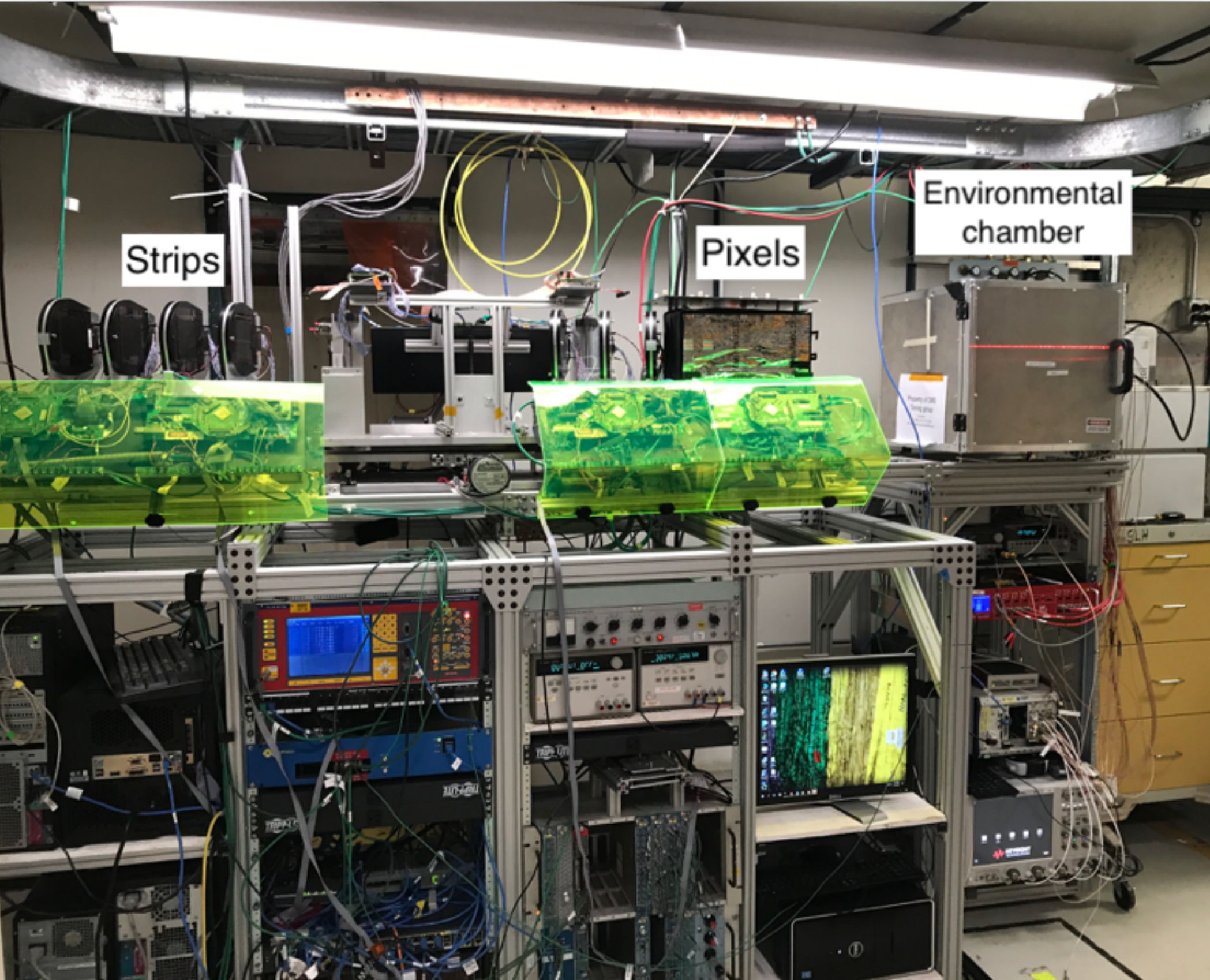}
    \caption{}
    \label{fig:FTBF}
  \end{subfigure}
  \begin{subfigure}[b]{0.4\linewidth}
    \begin{subfigure}[b]{\linewidth}
      \center \includegraphics[width=\linewidth, height=3cm, keepaspectratio]{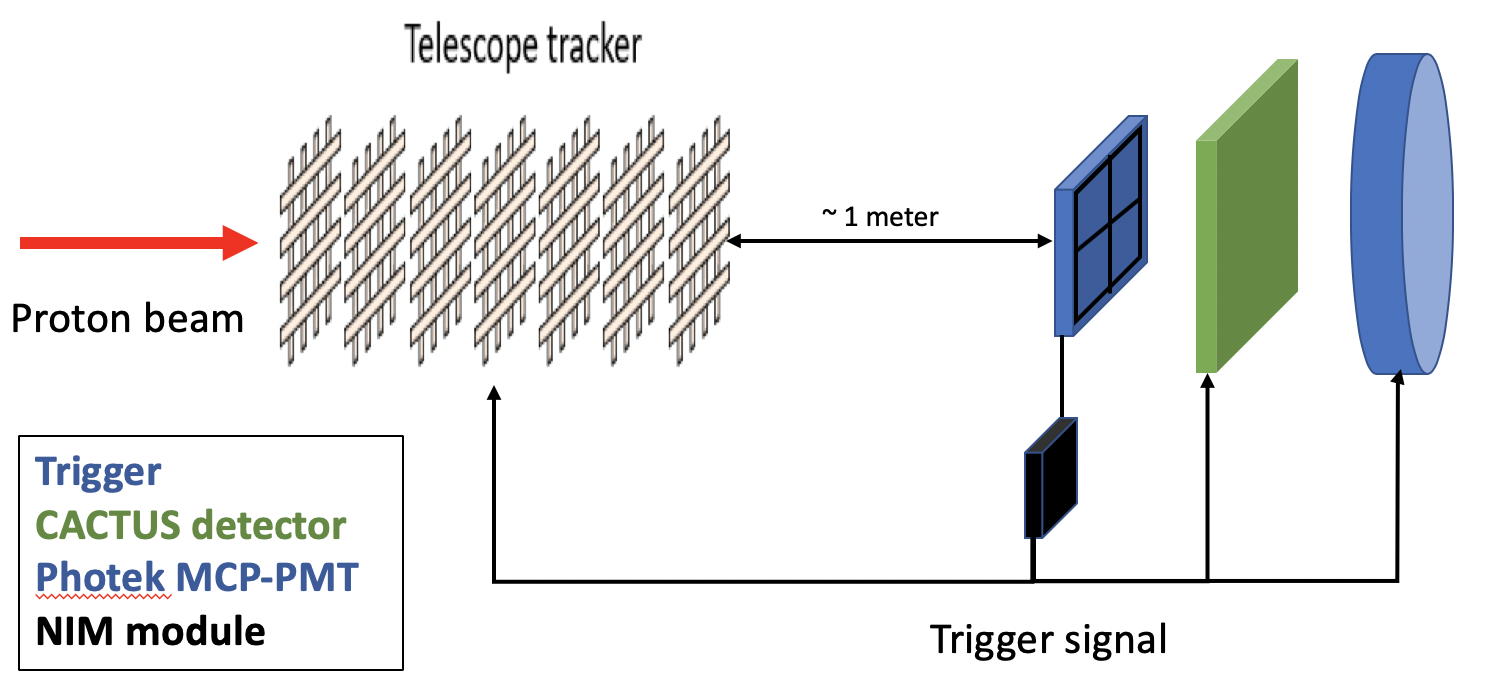}
      \caption{}
      \label{fig:FTBF_beam}
    \end{subfigure}
    \begin{subfigure}[b]{\linewidth}
      \center \includegraphics[width=\linewidth, height=3cm, keepaspectratio]{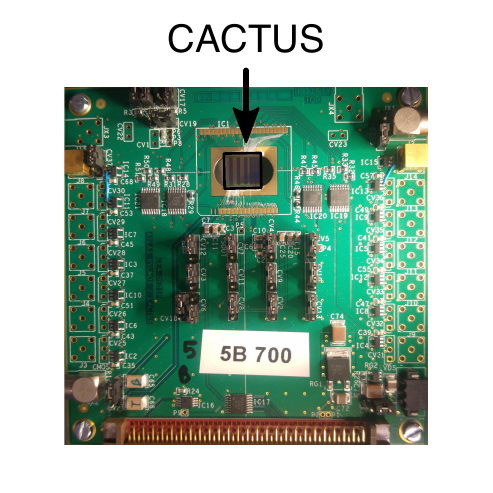}
      \caption{}
      \label{fig:CACTUS_BOARD:b}
    \end{subfigure}
  \end{subfigure}
\caption{a) A photo of the CACTUS characterization experimental station and the telescope tracker at FTBF, b) a schematic diagram of the test beam setup at FTBF, and c) a photo of the CACTUS board.}
\label{fig:TB_setup}
\end{figure}

The prototype CACTUS sensor was mounted on a remotely operated motorized stage, placed inside an environmental chamber with controlled temperature and humidity, shown in figure~\ref{fig:FTBF}. 
A schematic diagram of the experimental setup are shown in figure~\ref{fig:FTBF_beam}, which presents the arrangement of CACTUS sensor with respect to the telescope and triggers. The DUT board used in test beam measurements to characterize the CACTUS sensor is shown in figure~\ref{fig:CACTUS_BOARD:b}. The relative alignment was mechanically constrained to be around one mm, and we estimate the slant angle to be less than 5~degrees. The FTBF is equipped  with two silicon telescopes aligned along the beam
line and configured to operate synchronously. It has a pixel telescope
assembled from eight planes and a telescope with strip modules made up of
fourteen detector planes. Each microstrip plane consists of 639~microstrips,
each \SI{60}{\micro m} wide, placed orthogonal to each other, and the pixel telescope's cell sizes are \SI[product-units = power]{100 x 150}{\micro m}. The strip telescope increases the coverage of the pixel telescope and improves its tracking performance. The Data Acquisition (DAQ) hardware is based on the CAPTAN (Compact And Programmable daTa Acquisition Node) system developed at Fermilab. The CAPTAN is a flexible and versatile data acquisition system designed to meet the readout and control demands of a variety of pixel and strip detectors for high energy physics applications. 

\phantomsection
\label{MCP-PMT} A Photek 240~micro-channel plate (MCP-PMT) detector was placed furthest downstream behind the CACTUS sensor, inside the environmental chamber, and provided a very precise reference timestamp. Its precision has been previously measured to be less than \SI{10}{ps}~\cite{RONZHIN2015288}. The trigger to the telescope, the CACTUS sensor, and to the DAQ system was provided by a scintillator coupled to a PMT. The DAQ system is based on a Keysight MSOX92004A oscilloscope, which provides digitized waveforms sampled at \SI{40}{GS/s}. At the control room PC, the data from each CAPTAN node are saved in a binary file for each Run. Data from the CACTUS, the silicon telescope, and Photek were merged offline by matching the trigger counters of each system.

\subsection{Test beam results}
\label{sec:analysisresults}

\subsubsection{Detection efficiency}
\label{sec:efficiency}
Several adjacent pixels of \SI[product-units = power]{1 x 0.5}{\mm} in the middle of a CACTUS version~B sensor thinned to \SI{200}{\micro\metre} and backside polarized to HV~$=\SI{-300}{V}$ were tested during the testbeam.

\begin{figure}[h]
  \begin{subfigure}[b]{0.5\linewidth}
    \includegraphics[width=\linewidth]{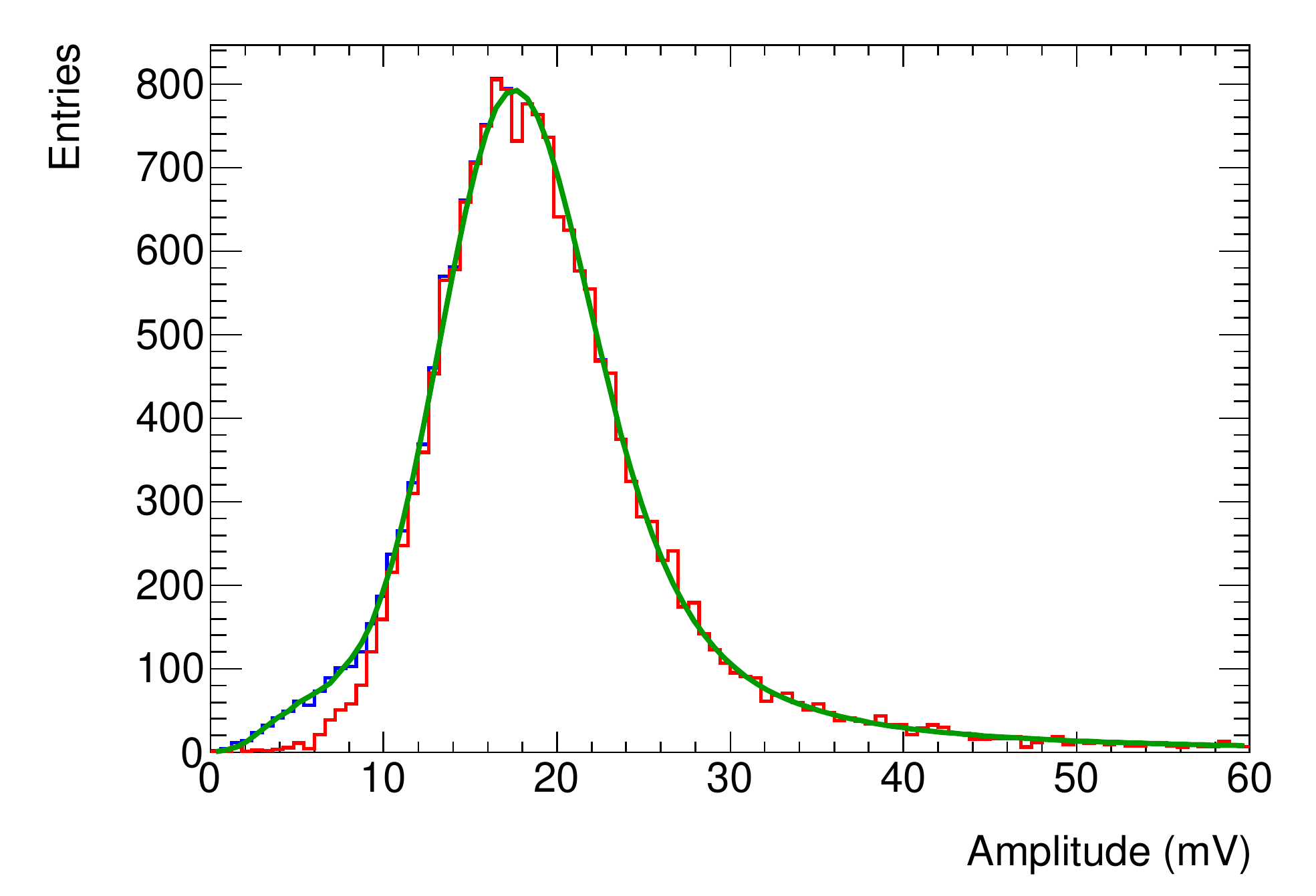}
    \caption{}
    \label{subfig-testbeam-effic-a}
  \end{subfigure}
  \begin{subfigure}[b]{0.5\linewidth}
    \includegraphics[width=\linewidth]{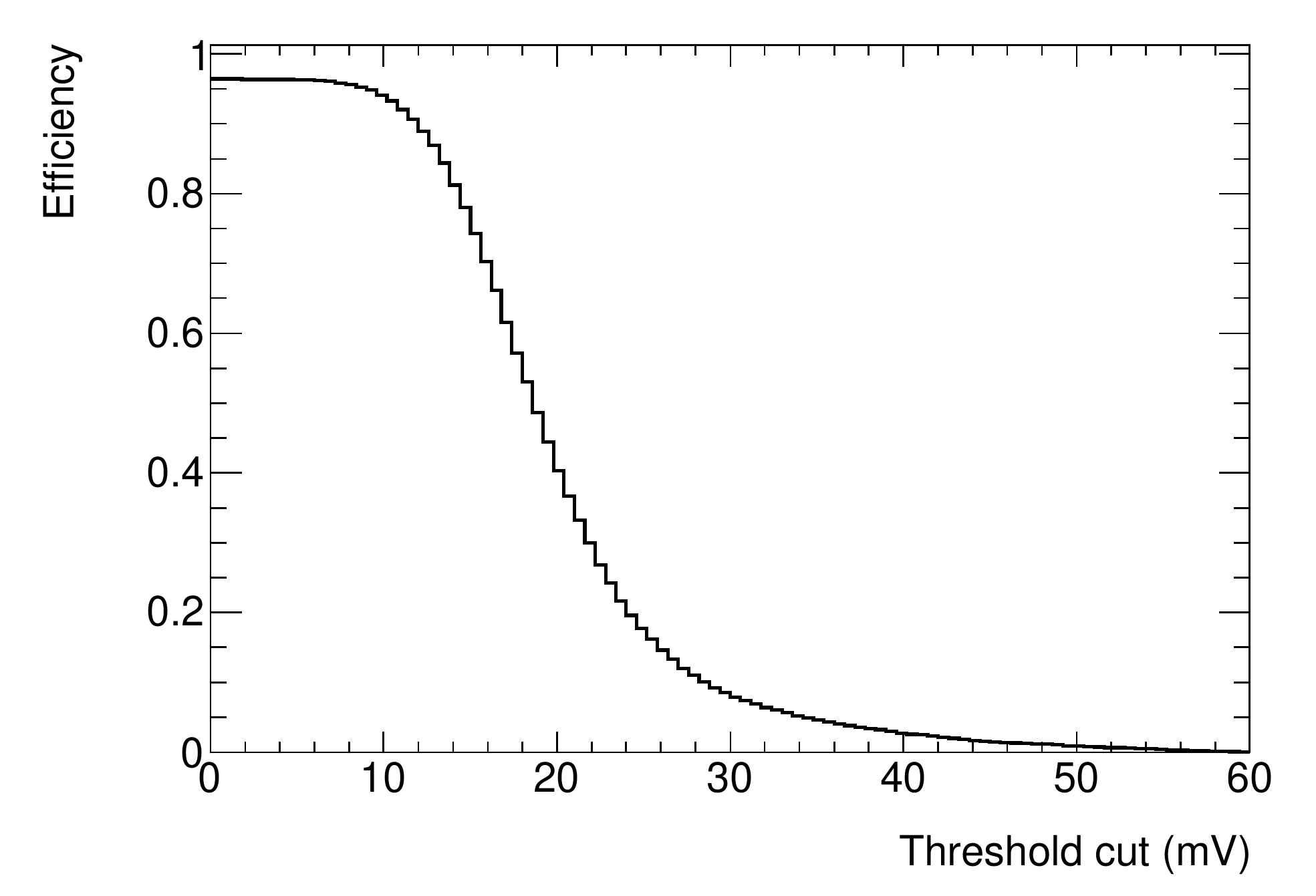}
    \caption{}
    \label{subfig-testbeam-effic-b}
  \end{subfigure}
\caption{a) Analog signal amplitudes measured on a pixel close to the center of the sensor. The data distribution is shown in blue while the noise-subtracted signal contribution is shown in red. The distribution is fitted to the sum of a Landau distribution convoluted with a Gaussian and a noise contribution. b) Efficiency as a function of the amplitude threshold in \si{\milli\volt}.}
\label{fig-testbeam-effic}
\end{figure}

The analog signal amplitude, defined as the height of the signal pulse in \si{\milli\volt} is shown in figure~\ref{subfig-testbeam-effic-a} for events where the telescope track crosses the pixel area. Only one pixel was read out at a time. A \SI{50}{\micro\metre} margin on the pixel's border was excluded to avoid possible bias due to the telescope tracking resolution. This distribution is fitted to the sum of the signal and noise distributions, with their relative normalization floated. The signal amplitude distribution is modeled by a convolution of a Landau distribution and a Gaussian.
A template for the noise amplitude distribution has been obtained from data using events where the telescope track points
outside the pixel area. The blue histogram shows the total data, while the red distribution shows the signal contribution after
statistical subtraction of the fitted noise distribution. The fit is shown in green, with the most probable value (MPV) measured at about  \SI{16.2}{\milli\volt} and a smearing (signal shape distorted by the noise) of about \SI{3.7}{\milli\volt}. The efficiency as a function of the threshold is deduced from the cumulative
distribution of the signal and shown in figure~\ref{subfig-testbeam-effic-b}. The slight \SI{3}{\percent} inefficiency that can be seen for a threshold of about \SI{10}{\milli\volt} is due to multiple
scattering after the telescope tracker, shifting the corresponding tracks out of the
pixel surface. The ratio between signal and background is estimated to be 5.5
for a threshold of \SI{10}{\milli\volt}.

\subsubsection{Response uniformity}
\label{sec:uniformity}
The analog signal uniformity has been evaluated by fitting the edges of the hit map
of the pixels to an \emph{erfc} function. A hit is defined as a signal in the pixel
with a maximum amplitude higher than \SI{20}{mV}, in time coincidence
with the test beam tracker. From the fits, the position
of the active edges have been extracted for the three pixels measured in the data. The position
of the active edge is defined as the position where the \emph{erfc} function reaches half of its asymptotic value.

\begin{figure}[h]
\center \includegraphics[height=5.2cm, keepaspectratio]{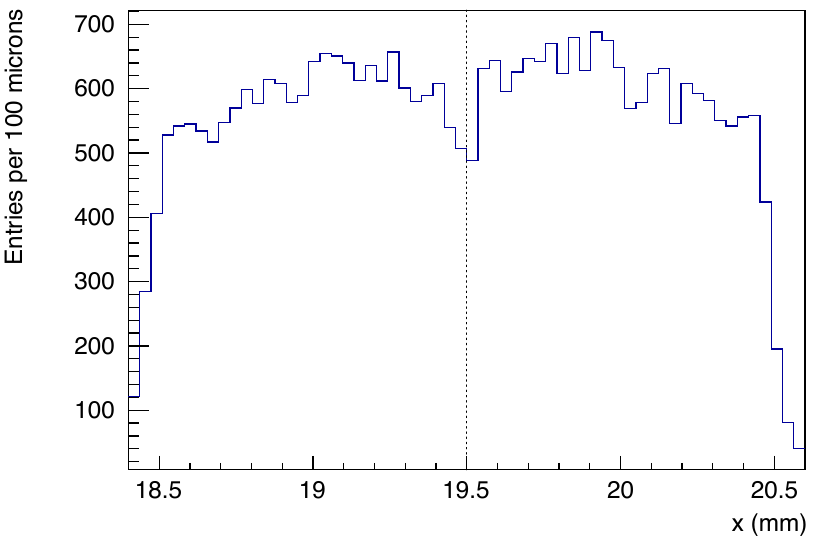}
\caption{Hit map projection on the $x$ axis, for two adjacent pixels. The vertical line indicates the boundary between two adjacent pixels.}
\label{uniformity:hitmap}
\end{figure}

Figure~\ref{uniformity:hitmap} shows a projection of the hit maps along the $x$ axis, superimposing hit maps obtained from two adjacent pixels. Only one pixel was read at a time. It can clearly be seen that there is a small hit rate reduction at the
boundary between two adjacent pixels. From the active edge position obtained from the fits, we observed a small inactive lane of \SI{20}{\micro\m} width in the $x$ direction, where the pixel pitch is \SI{1000}{\micro\m}, and of \SI{10}{\micro\m} in the $y$ direction, where the pixel pitch is \SI{500}{\micro\m}. As
a cross-check, the pixel sizes in $x$ and $y$ have also been measured and values that are consistent with \SI{1000}{\micro\m} and \SI{500}{\micro\m} were found.

The signal shape uniformity has also been checked as a function of the particle impact point position on the pixel. The pixel surfaces in strips of \SI{100}{\micro m} wide, in both directions, have been binned and
the average of all the pulse shapes corresponding to
particles impacting the pixel in each of these strips have been computed. 
%A small \SI{10}{\percent} variation of the amplitude with the pixel position was observed. However, normalized pulse shapes computed with the same procedure, but normalizing each pulse to its amplitude, did not show any difference, indicating that the shape itself is not correlated to the position.
A small \SI{10}{\percent} variation of the amplitude as a function of the track impact point position within the pixel was observed. However, normalized pulse shapes computed with the same procedure, but normalizing each pulse to its amplitude, did not show any difference. This indicates that the shape itself is not correlated to the track impact point position.

In addition to the same procedure, using five different bins in amplitude chosen to have the same number of events for all bins, it has been checked whether the pulses had the same shape for all amplitudes. No difference has been found.
This implies that the amplitude differences can be corrected within one pixel by implementing a simple TOT correction.

\begin{figure}[h!]
\includegraphics[width=\linewidth]{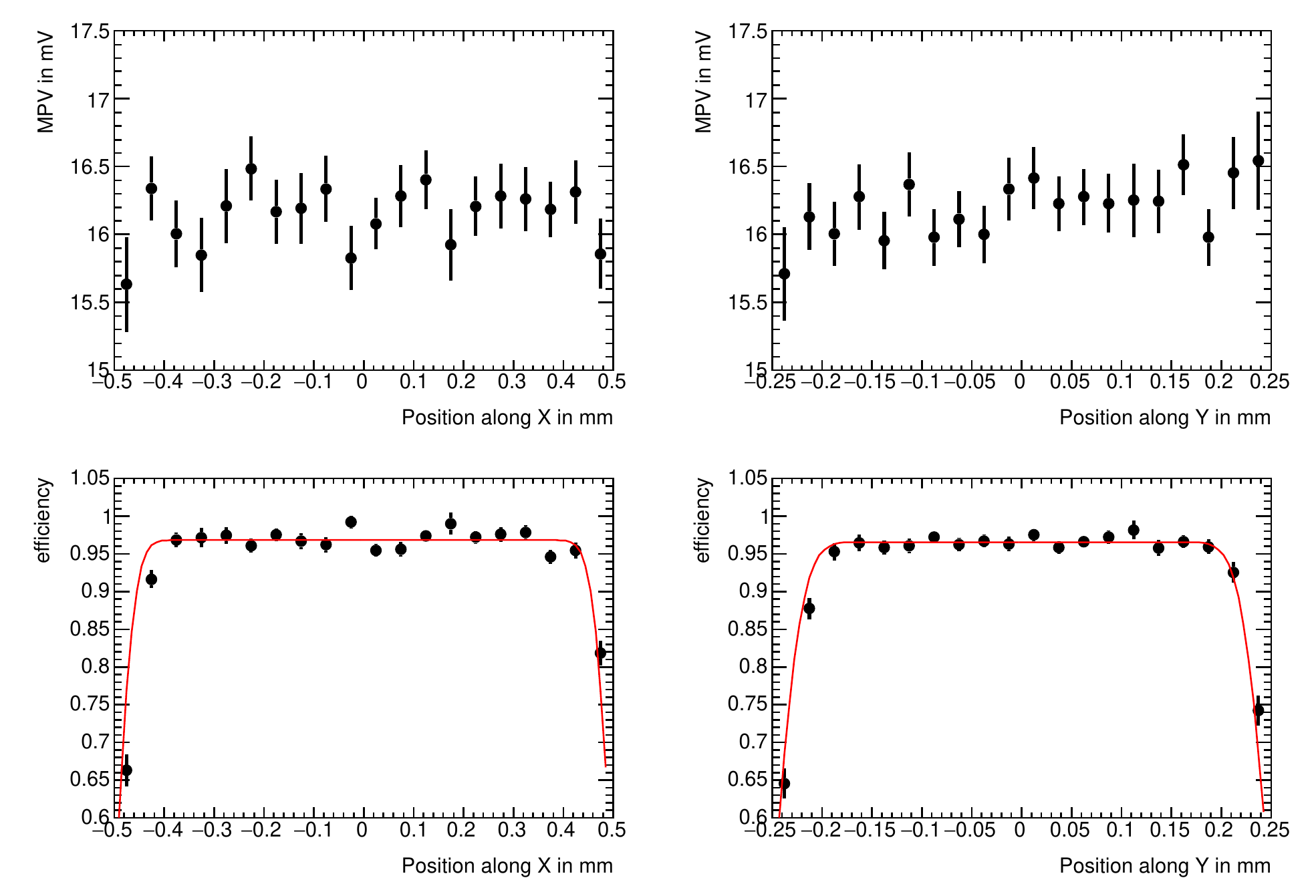}
\caption{MPV and efficiencies determined in slices of X or Y. For the efficiencies, the red curves show the expected efficiency when a \SI{97}{\percent} plateau efficiency is convoluted with a \SI{25}{\micro m} telescope resolution.}
\label{fig-testbeam-uniformity}
\end{figure}

Finally, the uniformity has also been checked by reproducing the fit on the amplitude distributions described in section~\ref{sec:efficiency} for several slices of X or Y of the tested pixel (excluding the \SI{50}{\micro\metre} margin in the other dimension). Figure~\ref{fig-testbeam-uniformity} shows the MPV of the Landau and the efficiency as a function of X or Y.

\subsubsection{Timing performance}
\label{sec:timing}
% FG : Proposal for this section
%\iftrue % true = in use
Test beam data have been used to evaluate the timing resolution of the sensor with MIPs. Due to the lower than expected signal to noise ratio, it was expected that the time resolution would be worse than \SI{180}{\pico\second} for that thickness and that it would be needed to cut strongly on the amplitude to obtain reasonable signals. Furthermore, the analog monitoring output is much slower than the analog front-end signal and was not intended to be used for timing studies (see section~\ref{sec:architecture}). Only one pixel was read out, which explains the efficiency drop visible at the edge of
the pixel.

\begin{figure}[h!]
\begin{subfigure}[b]{0.49\linewidth}
\includegraphics[width=\linewidth]{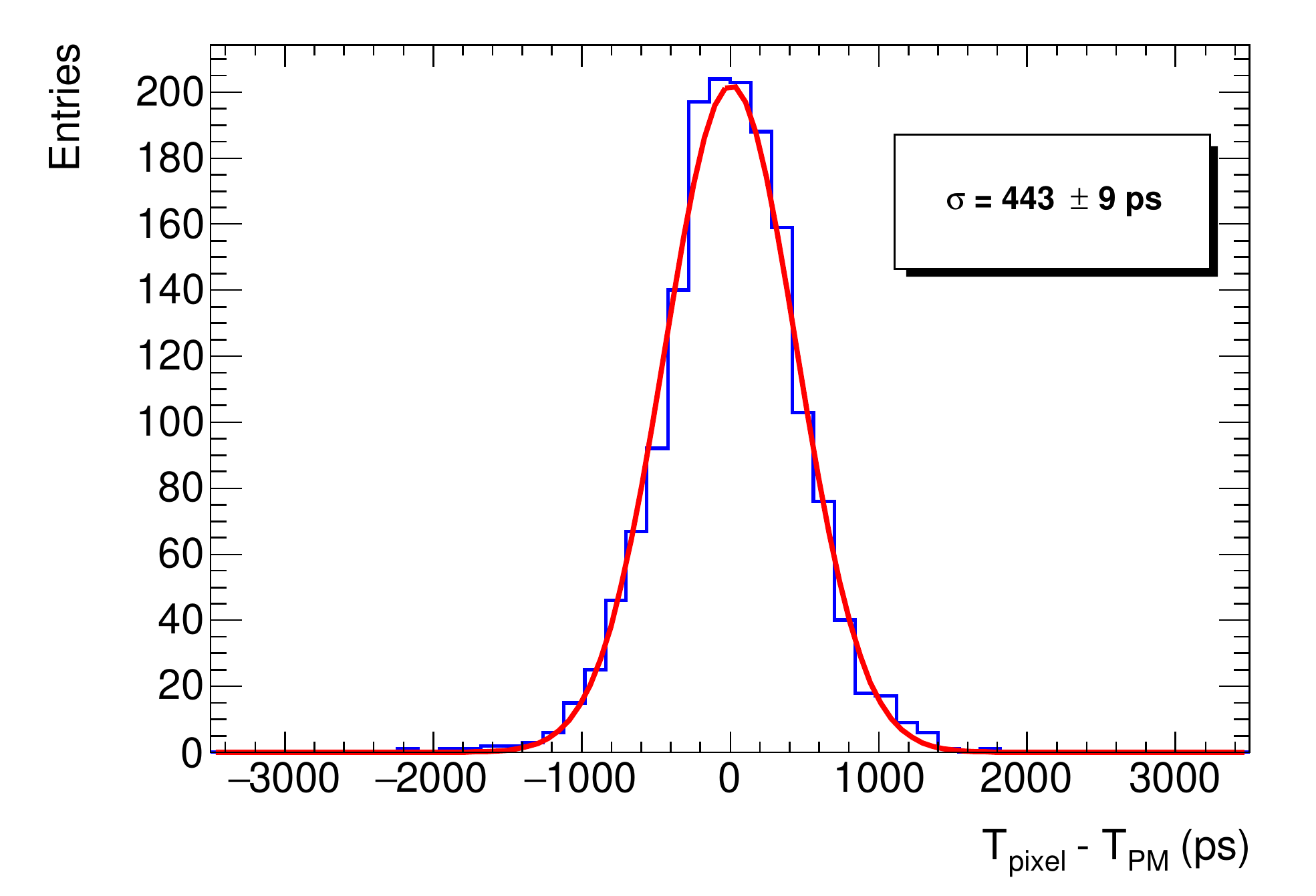}
\caption{}
\end{subfigure}
\begin{subfigure}[b]{0.49\linewidth}
\includegraphics[width=\linewidth]{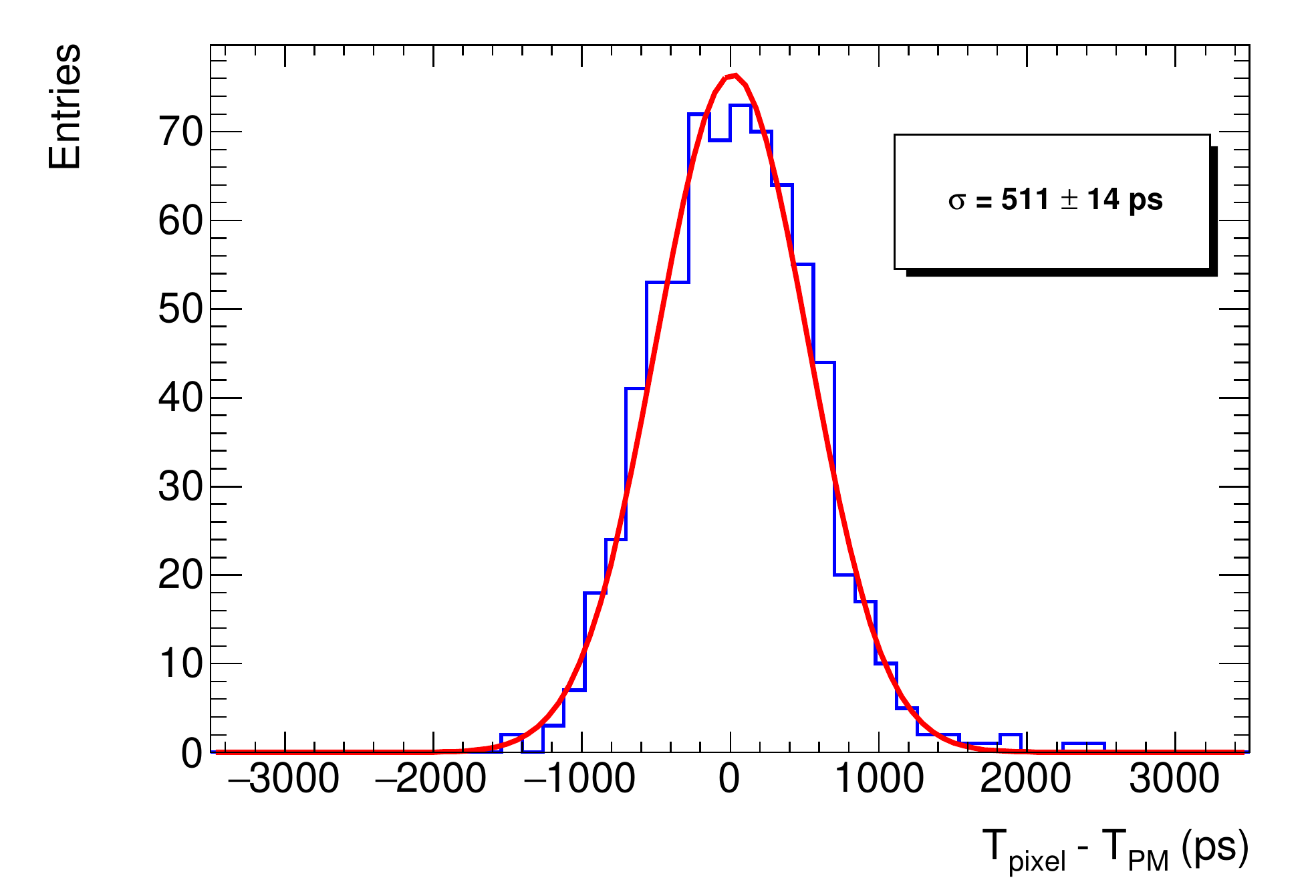}
\caption{}
\end{subfigure}
\caption{Time differences between sensor and PMT in \si{\pico\second} with a) analog monitoring output (threshold~$\simeq\SI{1.9}{MIPs}$) and b) digital output (threshold~$\simeq\SI{0.3}{MIPs}$).}
\label{fig:testbeam-timing}
\end{figure}

The signal measured from the sensor is compared to the signal from the Photek 240 MCP-PMT to obtain time measurements. For the analog readout, a quadratic time walk correction is obtained by measuring the time difference as a function of the amplitude. No time walk correction has been applied for the digital readout due to low statistics. The arrival time differences between the sensor and the PMT are shown for the analog output and the digital output in figure~\ref{fig:testbeam-timing}.

To minimize the distortion of the signal shape by the noise, which significantly degrades the time resolution at small signal amplitudes, we require
the signal amplitude to be larger than \SI{30}{\milli\volt} for analog data, corresponding to a threshold of $\simeq \SI{1.9}{MIPs}$. 
A lower threshold was used for the test beam digital runs ($\simeq \SI{0.3}{MIPs}$).

% YD : A similar cut was applied for the test beam digital runs ($\simeq \SI{0.4}{MIPs}$) leading to an equivalent efficiency for the two data sets ($< \SI{10}{\percent}$). 

The timing results from the MIPs are consistent with the results obtained with the $^{90}Sr$ source (figure~\ref{fig:testbeam-timing}).
%\fi

% FG : previous section for easy fallback
% SX: OK, i did not read and correct this version because the above version seemed better. 
%\iffalse
%Test-beam data have been used to evaluate the timing resolution of the sensor with MIPs. Due to the lower than expected signal to noise ratio, it was expected that the time resolution would be worse than \SI{100}{\pico\second}, and that it would be needed to cut strongly on the amplitude to get useable signals.

%To minimize the distortion of the signal shape by the noise which degrades
%significantly the time resolution of smaller amplitudes, we apply a cut
%at \SI{30}{\milli\volt} for
%analog data (which corresponds to a threshold of $\simeq \SI{1.9}{MIPs}$, to a bit less than \SI{10}{\percent} efficiency
%and is similar to the efficiency of the digital readout runs).
%The time measured by the sensor is compared to a reference PMT (of about \SI{10}{\pico\second} time resolution). For analog readout, a quadratic time walk is determined by making the difference amplitude-independent.
%The time differences between the sensor and the PMT are shown in figure~\ref{fig-testbeam-timing}.
%\fi

\section{Conclusions and perspectives}
\label{sec:conclusion}
A monolithic timing sensor prototype in a radiation hard CMOS \SI{150}{\nano\metre} process has been designed and tested. The test results are promising but some problems have been observed with this first CACTUS prototype. During the tests with radioactive sources, the observed analog signal to noise ratio has been found to be lower than expected from simulations and previous tracking prototypes as explained in section~\ref{subsec:S2N}. This observation has been confirmed with MIPs at test-beam. This effect is likely due to the in-pixel power metal rails, which increase significantly the capacitance of the charge collection diode. This problem will be overcome in future designs by modifying the pixel layout, mainly putting the front-end electronics outside of the charge collecting diode.

Other parameters of the chip have been shown to be very promising. The measured break-down voltages are high enough to deplete more than \SI{200}{\micro\metre}. Thinning and backside processing of fabricated wafers have been done reliably and successfully. The charge collection uniformity inside the pixel has been checked with MIPs, and no significant dead areas have been observed. 

A lower than expected signal to noise ratio presently limits the detection efficiency and the timing resolution of the detector. Nevertheless, the excellent global performance of the chip and our understanding of the observed imperfections call for a new design iteration of the CACTUS concept, and is already well under way.

\acknowledgments
We acknowledge the technical and financial support of IRFU-CEA for the design, characterization and tests done in lab for the CACTUS sensor. The experiments performed at the FTBF were supported by the U.S. Department of Energy. We wish to thank the Fermilab Test Beam Facility personnel for the continuous support they provide us. 

The testbeam measurements were performed using the resources of the Fermi National Accelerator Laboratory, a U.S. Department of Energy, Office of Science, HEP User Facility. Fermilab is managed by Fermi Research Alliance, LLC (FRA), acting under Contract No. DE-AC02-07CH11359. This work has also been supported by funding from the California Institute of Technology High Energy Physics under Contract DE-SC0011925 with the U.S. Department of Energy. 

%% FG : comment "note added"
%% \paragraph{Note added.} This is also a good position for notes added
%% after the paper has been written.


% BIBTEX + JINST style
\iftrue % true = in use
  \bibliographystyle{JHEP}
  \bibliography{jinst-CACTUS}

\providecommand{\href}[2]{#2}\begingroup\raggedright\begin{thebibliography}{10}

\bibitem{Allaire1}
C.~Allaire, J.~Benitez, M.~Bomben, G.~Calderini, M.~Carulla, E.~Cavallaro
  et~al., \emph{Beam test measurements of low gain avalanche detector single
  pads and arrays for the {ATLAS} high granularity timing detector},
  \href{http://dx.doi.org/10.1088/1748-0221/13/06/p06017}{\emph{JINST}
  {\bfseries 13} (Jun, 2018) P06017}.

\bibitem{CMS:2667167}
{CMS Collaboration}, \emph{{A MIP Timing Detector for the CMS Phase-2
  Upgrade}}, {\emph{Technical Design Report} {\bfseries
  \href{https://cds.cern.ch/record/2667167}{CMS-TDR-020}} (2019) }.

\bibitem{APRESYAN2018158}
A.~Apresyan, S.~Xie, C.~Pena, R.~Arcidiacono, N.~Cartiglia, M.~Carulla et~al.,
  \emph{Studies of uniformity of $50\mu$m low-gain avalanche detectors at the
  fermilab test beam},
  \href{http://dx.doi.org/https://doi.org/10.1016/j.nima.2018.03.074}{\emph{Nucl.
  Instrum. Meth. A} {\bfseries 895} (2018) 158 -- 172}.

\bibitem{Pieric1}
I.~Peri\'{c}, \emph{A novel monolithic pixelated particle detector implemented
  in high-voltage {CMOS} technology},
  \href{http://dx.doi.org/https://doi.org/10.1016/j.nima.2007.07.115}{\emph{Nucl.
  Instrum. Meth. A} {\bfseries 582} (Dec, 2007) 876 -- 885}.

\bibitem{Chen1}
Z.~Chen, M.~Barbero, P.~Barrillon, C.~Bespin, S.~Bhat, P.~Breugnon et~al.,
  \emph{Test results of irradiated {CMOS} pixel circuits in {150 nm CMOS}
  technology for the atlas inner tracker upgrade},
  \href{http://dx.doi.org/10.22323/1.343.0156}{\emph{PoS} (Jul, 2019) 156}.

\bibitem{Barbero1}
M.~Barbero, P.~Barrillon, C.~Bespin, S.~Bhat, P.~Breugnon, I.~Caicedo et~al.,
  \emph{Radiation hard {DMAPS} pixel sensors in {150nm CMOS} technology for
  operation at {LHC}}, {\emph{arXiv} (2019) },
  [\href{https://arxiv.org/abs/1911.01119}{{\ttfamily 1911.01119}}].

\bibitem{Iacobucci1}
G.~Iacobucci, R.~Cardarelli, S.~D{\'{e}}bieux, F.~D. Bello, Y.~Favre,
  D.~Hayakawa et~al., \emph{A 50~ps resolution monolithic active pixel sensor
  without internal gain in {SiGe} {BiCMOS} technology},
  \href{http://dx.doi.org/10.1088/1748-0221/14/11/p11008}{\emph{JINST}
  {\bfseries 14} (Nov, 2019) P11008}.

\bibitem{Riegler_2017}
W.~Riegler and G.~A. Rinella, \emph{Time resolution of silicon pixel sensors},
  \href{http://dx.doi.org/10.1088/1748-0221/12/11/p11017}{\emph{JINST}
  {\bfseries 12} (Nov, 2017) P11017}.

\bibitem{Degerli1}
Y.~Degerli, S.~Godiot, F.~Guilloux, T.~Hemperek, H.~Kr\"{u}ger, M.~Lachkar
  et~al., \emph{Pixel architectures in a {HV}-{CMOS} process for the {ATLAS}
  inner detector upgrade},
  \href{http://dx.doi.org/10.1088/1748-0221/11/12/c12064}{\emph{JINST}
  {\bfseries 11} (Dec, 2016) C12064}.

\bibitem{Degerli2}
Y.~{Degerli}, F.~{Balli}, M.~{Barbero}, S.~{Bhat}, P.~{Breugnon}, Z.~{Chen}
  et~al., \emph{Characterization of a new {HV/HR CMOS} sensor in {LF 150 nm}
  process for the {ATLAS} inner tracker upgrade},
  \href{http://dx.doi.org/10.1109/NSSMIC.2017.8532967}{\emph{IEEE NSS Conf.
  Rec.} (Oct, 2017) 1--3}.

\bibitem{Wang1}
T.~Wang, P.~Rymaszewski, M.~Barbero, Y.~Degerli, S.~Godiot, F.~Guilloux et~al.,
  \emph{Development of a depleted monolithic {CMOS} sensor in a 150 nm {CMOS}
  technology for the {ATLAS} inner tracker upgrade},
  \href{http://dx.doi.org/10.1088/1748-0221/12/01/c01039}{\emph{JINST}
  {\bfseries 12} (jan, 2017) C01039}.

\bibitem{Synopsys}
``Synopsys {TCAD}.'' \url{https://www.synopsys.com/silicon/tcad.html}.

\bibitem{Guilloux1}
F.~Guilloux, F.~Balli, Y.~Degerli, M.~Elhosni, C.~Guyot, T.~Hemperek et~al.,
  \emph{{CAcT$\mu$S: High-Voltage {CMOS} Monolithic Active Pixel Sensor for
  tracking and time tagging of charged particles}},
  \href{http://dx.doi.org/https://doi.org/10.22323/1.313.0023}{\emph{PoS} (Sep,
  2017) 023}.

\bibitem{WC}
D.~{Breton}, E.~{Delagnes}, J.~{Maalmi} and P.~{Rusquart}, \emph{{The
  WaveCatcher family of SCA-based 12-bit 3.2-GS/s fast digitizers}},
  \href{http://dx.doi.org/10.1109/RTC.2014.7097545}{\emph{IEEE NPSS Real Time
  Conf. Rec.} (May, 2014) 1--8}.

\bibitem{Caicedo1}
I.~Caicedo, M.~Barbero, P.~Barrillon, I.~Berdalovic, S.~Bhat, C.~Bespin et~al.,
  \emph{{The {Monopix} chips: Depleted monolithic active pixel sensors with a
  column-drain read-out architecture for the {ATLAS} Inner Tracker upgrade}},
  \href{http://dx.doi.org/10.1088/1748-0221/14/06/C06006}{\emph{JINST}
  {\bfseries 14} (Feb, 2019) C06006},
  [\href{https://arxiv.org/abs/1902.03679}{{\ttfamily 1902.03679}}].

\bibitem{Mandic1}
I.~Mandi\'{c}, V.~Cindro, A.~Gori\v{s}ek, B.~Hiti, G.~Kramberger, M.~Miku\v{z}
  et~al., \emph{{Neutron irradiation test of depleted {CMOS} pixel detector
  prototypes}},
  \href{http://dx.doi.org/10.1088/1748-0221/12/02/P02021}{\emph{JINST}
  {\bfseries 12} (Jan, 2017) },
  [\href{https://arxiv.org/abs/1701.05033}{{\ttfamily 1701.05033}}].

\bibitem{FTBF}
``{Fermilab Test Beam Facility}.'' \url{https://ftbf.fnal.gov}.

\bibitem{RONZHIN2015288}
A.~Ronzhin, S.~Los, E.~Ramberg, A.~Apresyan, S.~Xie, M.~Spiropulu et~al.,
  \emph{Study of the timing performance of micro-channel plate photomultiplier
  for use as an active layer in a shower maximum detector},
  \href{http://dx.doi.org/https://doi.org/10.1016/j.nima.2015.06.006}{\emph{Nucl.
  Instrum. Meth. A} {\bfseries 795} (2015) 288 -- 292}.

\end{thebibliography}\endgroup


\begin{thebibliography}{99}
    \input{tex/OLD-biblio.tex}
  \end{thebibliography}
\fi

% FG : Previous biblio
\iffalse % false = not used
  
\fi

% Please avoid comments such as "For a review'', "For some examples",
% "and references therein" or move them in the text. In general,
% please leave only references in the bibliography and move all
% accessory text in footnotes.

% Also, please have only one work for each \bibitem.

\end{document}